# Privacy-Preserving Data Processing in Cloud : From Homomorphic Encryption to Federated Analytics

Gaurav Sarraf, Vibhor Pal

Independent Researcher, USA

ABSTRACT

Privacy-preserving data processing refers to the methods and models that allow computing and analyzing sensitive data with a guarantee of confidentiality. As cloud computing and applications that rely on data continue to expand, there is an increasing need to protect personal, financial and healthcare information. Conventional centralized data processing methods expose sensitive data to risk of breaches, compelling the need to use decentralized and secure data methods. This paper gives a detailed review of privacy-saving mechanisms in the cloud platform, such as statistical approaches like differential privacy and cryptographic solutions like homomorphic encryption. Federated analytics and federated learning, two distributed learning frameworks, are also discussed. Their principles, applications, benefits, and limitations are reviewed, with roles of use in the fields of healthcare, finance, IoT, and industrial cases. Comparative analyses measure trade-offs in security, efficiency, scalability, and accuracy, and investigations are done of emerging hybrid frameworks to provide better privacy protection. Critical issues, including computational overhead, privacy-utility trade-offs, standardization, adversarial threats, and cloud integration are also addressed. This review examines in detail the recent privacy-protecting approaches in cloud computation and offers scholars and practitioners crucial information on secure and effective solutions to data processing.

## Introduction

The issue of data privacy protection is a burning issue in the academic community as well as in business because data-driven technology is growing exponentially [1]. Data security is of the utmost importance due to the large amount of sensitive information that is created and processed every day. Data is typically pooled on central servers for conventional central storage and computation-based data processing solutions. However, transporting and storing data via this centralized system is likely to violate privacy. It is extremely dangerous for individuals and organizations when sensitive information is at risk of disclosure due to a compromise or attack on the central storage or transmission process.

Federated learning has become an increasingly popular approach to decentralized data processing to solve these privacy issues [2][3]. Federated learning can train models through the collaboration of multiple devices or entities, without raw data being shared, which mitigates to a certain degree the privacy risk. Nevertheless, federated learning can overcome certain privacy issues of centralized systems but has its vulnerabilities nonetheless. Transmission of model updates in a federated system may unwillingly reveal information about a local dataset even in a federated system. This has prompted researchers to focus on improving privacy protections for federated learning systems.

This is where homomorphic encryption (HE) has proven to be a useful technique for protecting privacy in federated learning. In HE, encrypted data can be subjected to computations without needing to be decrypted [4], and so sensitive data can be kept safe during the full data processing chain. HE is a top choice for federated learning privacy because of this property.

## Structure of the Paper

The paper is organized as follows: Section II provides the background and fundamentals, Section III presents privacy preserving methods, Section IV deals with collaborative and emerging frameworks, Section V deals with applications and comparative analysis, Section VI deals with literature, and Section VII makes conclusions and presents the main challenges.

## Background and Fundamentals

Cloud computing offers on-demand and scalable services of computing resources and storage services, which organizations can utilize to deal with large amount of data easily. Nonetheless, this centralized architecture poses a significant privacy risk, with sensitive data such as personal health records or financial information most likely being kept and processed on third-party servers. Some of the critical issues are unauthorized access, data breach, insider threat and misuse of personal information. The fact that regulatory compliance is another obstacle to the process and that frameworks like GDPR, HIPAA, and CCPA impose severe demands on how data is handled, stored, and shared further complicates the matter. Moreover, cloud multi-tenancies put more data at risk of leaking between customers, and offloading computation to third parties casts doubt on data confidentiality and control. Cloud service trust relies on resolving these privacy concerns, which in turn pave the way for the implementation of privacy-preserving measures like encryption, differential privacy, and federated analytics.

## Essential Characteristics of Cloud Computing

According to the NISO definition, Table I below outlines the five primary components of a cloud model [5].

TABLE I. CLOUD COMPUTING ESSENTIAL CHARACTERISTICS

| Techniques | Description |
|---|---|
| Homomorphic Encryption (HE) | The ability to directly process encrypted data yields results that are identical to those of plaintext operations upon decryption. Achieves privacy-preserving analytics, secure machine learning, and outsourced compute. There are two varieties of encryption: fully homomorphic and partially homomorphic. |
| Secure Multi-Party Computation (SMPC) | Collaboratively calculating a function using secret inputs allows many parties to remain anonymous. Oblivious Transfer, Secret Sharing, and Yao's Garbled Circuits are some of the most used protocols. Beneficial for federated computations and collaborative analytics. |

| Differential Privacy (DP) | Implements a method to conceal certain data points by adding controlled noise to query results or models. Includes Centralized DP (CDP) and Local DP (LDP). Applications include cloud-based analytics, machine learning, IoT, and healthcare. |
|---|---|
| Federated Analytics / Federated Learning (FA/FL) | All data stays within the local system; only updated models or aggregated results are transferred. Can be used in conjunction with HE or DP to improve privacy. Applications in healthcare, finance, and collaborative ML across organizations. |
| Trusted Execution Environments (TEE) | Secure enclaves built into hardware (such as Intel SGX and AMD SEV) encrypt information as it is processed. Ensures secure cloud processing without exposing plaintext to providers. |
| Data Anonymization & Pseudonymization | Methods like k-anonymity, l-diversity, and t-closeness can de-identify sensitive data. Implemented for the purpose of public dataset sharing and regulatory compliance. |
| Access Control & Identity Management | Access controls that are based on roles or attributes aim to restrict access to sensitive information to only permitted people. Cryptography and auditing are frequently used together. |
| Secure Data Aggregation | Combines multiple data sources securely, often used in IoT and sensor networks. Prevents exposure of individual contributions while computing global statistics. |

## Cloud Architectures

A "cloud architecture" is a description of the structure and layout of a cloud computing environment, including the allocation of resources, applications, and services across various cloud service providers. Cloud architecture is crucial for a cloud computing environment's scalability, reliability, and security. A key component to understanding and effectively utilizing cloud computing is realizing that it is fundamentally a simultaneous service [6]. The safe storage, protection from natural disasters, and easy and secure flow to and from the user are all essential requirements for valuable resources such as data, information, and knowledge.

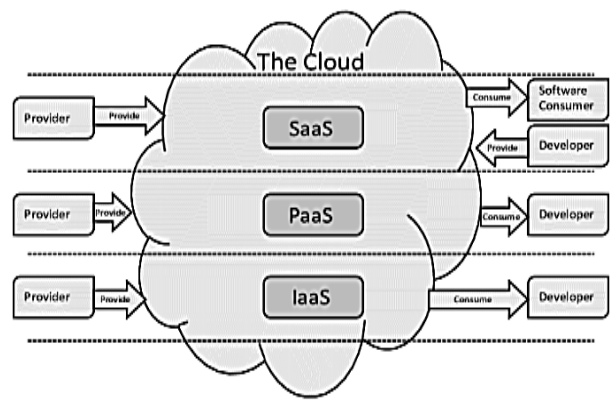

Figure 1. Cloud Computing Architecture

Computing, storage, database management, application services, and large-scale computing are just a few of the many IT processes that have benefited from cloud computing. Expertise in cloud computing relies heavily on its architecture (Figure 1), which describes the system's components, subsystems, and general organization [7]. Many distinct approaches exist for implementing cloud computing, and each has its own set of advantages and disadvantages. These designs provide advice on how to build and implement cloud-based application designs. In most cases, the following essential parts make up cloud architecture:

1) **Cloud Service Models**
   - **IaaS:** Providing online virtualization services such as cloud computing, storage, and networking.
   - **PaaS:** Enables users to focus on application creation, release, and oversight rather than the restrictions of the underlying infrastructure.
   - **SaaS:** Distributes software programs over the internet, relieving users of the burden of installing, maintaining, and monitoring these programs on a personal level.

2) **Deployment Models**
   - **Public Cloud:** The facilities are accessible to all individuals who wish to utilize them through the open internet.
   - **Private Cloud:** Private networks, typically used within an organization, provide for more control and personalization when providing services [8].
   - **Hybrid Cloud:** Facilitates app and data sharing by combining features of public and private clouds.

3) **Cloud Services**
   - **Identity and Access Management (IAM):** The ability to manage user identities and permissions effectively safeguards cloud resources from invasion.
   - **Database Services:** Managed database support: support of managed databases with secure and expandable data storage.

4) **Orchestration and Management**
   - **Automation:** Tools and resources for automating routine tasks, reducing human intervention, and increasing output capacity.
   - **Orchestration:** Ensuring easy communication through administration and management of several cloud services.

5) **Scalability and Elasticity**
   - **Scalability:** The capacity of a system to handle increased workload by incorporating additional resources is known as its scalability.
   - **Elasticity:** The ability to automatically modify resources to meet demands.

Cloud architecture is dynamic and adaptable, constantly evolving to accommodate new technologies. When designing their cloud infrastructures, businesses consider their own needs in terms of efficiency, effectiveness, cost-effectiveness, and compliance [9]. Remember that there isn't a common architecture or even fundamental design principles for cloud-based apps just yet. An existing objective of cloud computing is the establishment of a controlled and scalable cloud infrastructure. This design has to be scalable to variations in service levels and market-oriented so that it can respond to fluctuations in the demand for and supply of cloud resources.

## *Understanding Regulatory Compliance and Cloud Data Protection*

This regulatory compliance is a very important process in cloud data protection because it verifies that the organizations are in compliance with the laws and standards that have been put in place to safeguard privacy, security and integrity of the data. The increased concern over data breaches, cyberattacks and privacy invasions has increased regulatory pressure [10]. The non-compliance may lead to legal, financial and reputational losses to the organization which fails to adhere to the compliance standards. The fact that there are regulatory requirements peculiar to the protection of cloud data therefore makes it important that organizations, especially those operating in an industry that handles sensitive information, are aware of and meet these demands.

Regulatory compliance is the observance of legal, ethical and industry specific regulations concerning the gathering, storing, processing, and disseminating of information. Compliance, as it pertains to cloud computing, means that any data stored in the cloud is handled in line with all applicable rules and regulations. These laws can be regional, industrial in nature, and the nature of the data. Compliance also depends on carrying out required security controls, audit and documentation to show compliance with these regulations.

6) **Global Regulatory Landscape for Cloud Data Protection**

*The General Data Protection Regulation (GDPR)*

The purpose of the GDPR was to ensure the security of personal information and data belonging to EU residents. Data encryption, breach notification, and foreign data transfers are just a few of the many restrictions it places on businesses in relation to the collection, storage, and handling of consumers' personal information. Organizations must take the necessary organizational and technical steps to safeguard personal data in accordance with GDPR. This applies regardless of whether the data was kept on the cloud or not.

*The Health Insurance Portability and Accountability Act (HIPAA)*

Privacy and security regulations for PHI are set out by the HIPAA of the United States. The providers of healthcare, insurers, and cloud providers that work with PHI need to adhere to HIPAA standards, such as the necessity of PHI encryption and protection at rest and during transmission. Cloud vendors are expected to provide the security required to be compliant with HIPAA and healthcare organizations should make sure that the type of cloud services being utilized to process PHI is in compliance with the required security and privacy standards of HIPAA.

*The California Consumer Privacy Act (CCPA)*

The CCPA is a piece of state-level privacy legislation that gives Californians more control over their personal information. The ability to access, amend, or delete one's personal data as well as to decline its sale would all fall under these protections. The CCPA stipulates that cloud service providers which process data of California residents must adhere to the requirements of the document, which is transparency about data collection practices, data security, and the establishment of appropriate consent management procedures.

**Privacy-Preserving Techniques In Cloud Computing**

In cloud computing environments, the next section talks about some methods to guarantee privacy-preserving data processing:

*Homomorphic Encryption (HE)*

A cryptographic procedure known as HE allows one to compute values from encrypted data that are identical to those obtained from the plaintext by applying the same operations on the encrypted data. Use of this feature allows analysts to protect the privacy of sensitive information. Although the idea was initially proposed in 1978 by Rivest, Adleman, and Detrusors, it wasn't until the early 2000s that workable implementations started to emerge. The most recent development in this area was Gentry's 2009 FHE method, which laid the groundwork for later efforts. Examples of PHE include RSA and El Gamal; SHE allows for a limited number of operations; while FHE allows for unrestricted calculations. Some examples of SHE are RSA and El Gamal. Although it is secure and computationally intensive, FHE is impossible to use in large-scale and real-life applications.

7) **Applications of Homomorphic Encryption in Cloud Computing**

HE is used in cloud computing for the following reasons:

- **Secure Data Analytics:** HE facilitates the execution of cloud-based calculations on confidential data without the disclosure of that data itself, hence protecting the privacy of organizations [11][12].

- **Healthcare Data Protection:** Cloud computing allows for the secure storage and analysis of patient data, allowing for the execution of critical calculations in a way that is fully compliant with all applicable standards.
- **Privacy-Preserving Machine Learning:** Collaborative learning with secure data is made possible with HE when using encrypted datasets for training and inferring ML models (such support vector machines and linear regression).
- **Secure Multi-Party Computation (SMPC):** Computation of a function over private inputs by several parties without disclosure of those inputs guarantees data secrecy.
- **Data Sharing Across Organizations:** HE facilitates secure sharing and collaborative computation on sensitive data distributed across different entities.

8) **Challenges and Limitations of Homomorphic Encryption (HE)**

Here are some challenges and limitations of homomorphic are discussed below:

- **High Computational Overhead:** Encryption, decryption, and operations on ciphertexts are computationally intensive, causing performance bottlenecks.
- **Inefficiency in Time-Sensitive Applications:** HE is not effectively used in real-time or other latency-sensitive applications because of processing delays [13].
- **Complex Implementation Requirements:** Implementation of HE involves specialization in the mathematical principles of the technique and selection of parameters.
- **Limited Adoption:** Technical complexity has not enabled widespread use in industry and research.
- **Lack of Standardized Tools and Frameworks:** There are limited frameworks and practical tools that are used in the implementation of HE in practice.
- **Scalability Concerns:** Data-intensive or large-scale applications are highly challenged in regard to the computation and store overheads.

*Secure Multi-Party Computation (SMPC):*

SMPC protocols are used to guarantee privacy and integrity of computations, which consist of more than one participant. These protocols are also to be secure against the intrusion of adversaries who are interested in the privacy of the participants as well as the interference with the computation process [14][15]. Security models are therefore important in establishing the capabilities and intent of possible adversaries. In this section, discuss the technical aspects of SMPC, which includes the mathematical basis, algorithm implementation, and cryptographic protocols.

9) **Cryptographic Protocols in SMPC**

SMPC employs a number of cryptographic techniques to guarantee the confidentiality of each individual data input throughout the calculation process. Confidential information sharing, homomorphic encryption, and undetected transfer are all examples of such protocols.

- **Secret Sharing:** Allows breaking down an input into multiple shares which can be distributed to participants. Each share in itself might have no value, but when brought together they can recreate the starting point without revealing it to an individual.
- **Homomorphic Encryption:** Performs calculations using encrypted data, which results in an encrypted answer, which, on decryption, concurs with the outcomes produced on the plaintext [16]. This feature is characterized by ensuring that information is processed with security and encryption.
- **Oblivious Transfer:** Enforces the right of the recipient to privacy in the sense that one party is able to send messages to another without deciding whose information was received.

## 10) Privacy and Security In Secure Multi-Party Computation (SMPC)

Secure Multi-party Computation (SMPC) is a framework that gives a computation of functions on distributed data without revealing the input of the participant. This part discusses the privacy and security guarantees offered by SMPC, the type of threats it is able to withstand, and the limitations on data protection that it has.

SMPC guarantees the privacy of a party so that no party possesses more information than is needed regarding the outcome of the calculation. This can be done using cryptographic protocols, by which an calculation can be done without knowing specific inputs [17]. The privacy of any specific inputs is maintained in SMPC, as long as the majority of the participants do not collude, which can be usually assured assuming that a specific number of collaborating parties will be honest.

SMPC Security Guarantees: The ability to perform computations correctly and consistently in spite of the availability of aggressive elements is known as security in SMPC. SMPC protocols should be capable of enduring a number of hostile attacks, including manipulation and eavesdropping. They also assume that some of the participants also be tainted or act maliciously and ensure that the result calculated is correct.

### *Differential Privacy (DP)*

A mathematical concept known as differential privacy helps to safeguard the privacy of individuals and yet permits the analysis of data. It operates by adding appropriate level of noise into the data or the responses of a query into the data such that the presence (or absence) of one data point does not significantly influence the answer. Such an approach guarantees that one can hardly infer the data about one particular person based on the accumulated results providing high privacy levels. Differential privacy is important when private information like medical records or financial transactions needs to be looked into to learn more without revealing the identities of the people involved [18]. Figure 2 shows how differential privacy can be used to secure user data while still enabling data analysis. Collect sensitive user information from all sorts of locations, including hospitals and schools. Taking the data's sensitivity into account, and apply a certain function—the noise function—to the original data, transforming it in the process. In doing so, hope to preserve data utility while simultaneously protecting individuals' privacy. Without violating anyone's privacy, methods like DL are used to process the noisy data and derive useful insights.

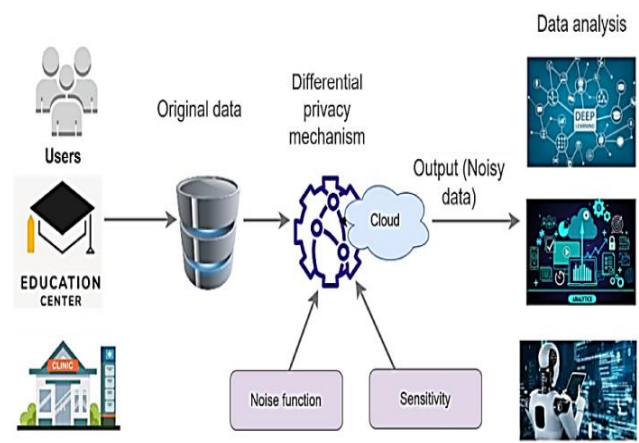

Figure 2. Differential Privacy Concept

Differential privacy is important because it provides robust and quantifiable privacy protections, which are becoming more important in the era of big data and advanced analytics. It is critical to protect individuals' privacy within these databases as organisations rely more and more on big datasets for innovation and decision-making [19]. Differential privacy safeguards data privacy by decreasing the likelihood of reidentification or the exposure of sensitive information, even in the event that an adversary has access to additional information. Ensuring compliance with strict privacy requirements, including the general data protection regulation, and keeping the trust between data suppliers and consumers both depend on this.

### Collaborative and Emerging Privacy Frameworks

This section covers Federated Analytics & Learning (FA/FL) and Emerging Trends, highlighting collaborative methods and next-generation frameworks for privacy-preserving cloud computing.

### *Federated Analytics & Federated Learning (FA/FL)*

Federated Analytics and Federated Learning is a paradigm shift in privacy preserving computation, whereby raw data is not sent to a central server, but intermediate results or model updates are sent to a central server. The model resolves the privacy concerns of centralized data aggregation, and especially applicable in such highly sensitive areas like healthcare and finance.

11) **Paradigm**

FA focuses on collaborative statistical analysis, while FL extends this to ML, enabling joint model training without direct data sharing.

12) **Secure Aggregation**

Cryptographic protocols, e.g., SMPC and HE [20], can be utilized to make sure that the individual contributions are obscured, and only the aggregated results can be seen by the server.

13) **Integration with DP & HE**

Differentiated privacy (DP) is frequently added to federated systems to reduce threats of model inversion and privacy of membership, and HE can be used to compute with encrypted updates. The overlaying methodology provides local and global privacy assurances.

14) **Real-World Deployments**

Google's Board is one of the earliest mass FL applications, as the company keeps user typing data locally and only shares model improvements [21]. In the case of FL, health consortia can train models that span across numerous hospitals to predict diseases without providing any patient information. Financial institutions have also started to use FL when it comes to credit scoring and fraud detection.

15) **Challenges**

In spite of its potential, FA/FL has been challenged by communication overhead, the system being highly heterogeneous, data imbalance (non-IID distributions) and susceptible to model poisoning or backdoor attacks. These problems are essential in order to solve real-world scalability.

### *Emerging Trends*

In addition to FA/FL, various new models and technologies are under development that can enhance privacy assurances in cloud-based data processing.

16) **Trusted Execution Environments (TEE)**

TEEs, including Intel SGX and AMD SEV, allow executing computation on plaintext data defined in isolated hardware enclaves, and provide almost native efficiency [22]. Although TEEs decrease the cryptographic overhead, they are susceptible to hardware attacks and side-channel attacks.

17) **Hybrid Approaches (HE + DP + FL)**

This mixture of methods is becoming increasingly popular in recent studies as a way of developing a compromise between security, scalability, and accuracy. An example is secure aggregation via HE, which is same-side, and decentralized training via FL, and noise inference attacks via DP. Such hybrid frameworks provide layered protection against a diverse set of threats.

**18) Quantum-Resistant Methods**

Classical encryption schemes may become obsolete as quantum computing continues to evolve. Cloud privacy-preserving methods are now using post-quantum cryptography based on lattice issues (e.g., Ring-LWE) to ensure the security of HE and SMPC protocols in the post-quantum era.

**19) Edge-Cloud Collaborative Frameworks**

Emerging architectures combine edge computing with cloud analytics, allowing sensitive data to be pre-processed and anonymized at the edge (e.g., IoT devices, mobile phones) before being aggregated in the cloud [23]. Applications involving real-time data, such as smart cities, autonomous vehicles, and industrial IoT, benefit from this design's lack of delay and raw data exposure.

**Applications and Future Directions**

The healthcare, financial, IoT, and industrial sectors can all benefit from privacy-preserving data processing, which allows for the safe and efficient exchange of data. A wide range of approaches provide varying degrees of security, scalability, accuracy, and efficiency; these include hybrid methods, federated learning, differential privacy, safe multiparty computation, and homomorphic encryption, among many others [24]. While these approaches support applications like secure EHR exchange, privacy-aware financial analytics, and predictive maintenance, challenges remain in scalability, privacy–utility trade-offs, standardization, adversarial threats, and cloud integration. The following section presents a discussion on the applications, comparative analysis, and associated challenges of privacy-preserving data processing techniques:

*Applications in Healthcare, Finance, IoT, and Industry*

Here, some of the applications related to healthcare, finance, IoT and Industry are discussed below:

- **Healthcare:** Enables secure exchange of electronic health records (EHRs) across institutions while maintaining patient confidentiality [25]. It also allows privacy-preserving genomic analysis for precision medicine and secure remote monitoring via wearable and sensor-based devices.
- **Finance:** Achieves compliance with data privacy requirements while facilitating the identification of fraud and the assessment of credit risk using encrypted transaction data. Favors privacy-conscious customer profiling, personalised financial services and safe blockchain-based payment systems.
- **Internet of Things (IoT):** Secures vulnerable information of smart homes, connected cars, and wearables. Supports safe information transfer within healthcare IoT and provides privacy within industrial sensor systems and can support a range of applications such as predictive maintenance.
- **Industry Case Studies:** Federated learning has been applied in manufacturing to make predictions of maintenance and quality in products without sharing of raw data. Privacy-sensitive analytics are used in supply chains to optimize logistics, tracking and demand forecasting. Methodologies assist in the safe management of smart grids and privacy-conscious demand-response processes in the energy sector.

*Comparative Analysis: Taxonomy of Techniques*

A variety of methods exist for processing personal data, each with its own advantages and disadvantages. The taxonomy of these techniques can be broadly categorized into cryptographic methods, statistical privacy approaches, distributed learning methods, and hybrid solutions [26]. Secure multiparty computing and HE are two examples of cryptographic algorithms that provide high levels of security with a hefty computational cost. Statistical approaches, including differential privacy, provide tenable privacy levels with relatively high efficiency but may impact data accuracy. Distributed learning methods, particularly federated learning, enable scalable analytics by keeping raw data decentralized, though they may be susceptible to adversarial attacks. The scalability, precision, efficiency, and security that hybrid systems achieve by combining numerous techniques make them ideal for complex real-world applications. The main features and disadvantages of these methods are summarised in Table II.

TABLE II. COMPARATIVE ANALYSIS: TAXONOMY OF TECHNIQUES

| Techniques | Security | Efficiency | Scalability | Accuracy | Key Synergies |
| --- | --- | --- | --- | --- | --- |
| Homomorphic Encryption (HE) | Very High | Low (computationally expensive) | Low–Medium | High | Effective when integrated with federated learning for secure training |
| Secure Multiparty Computation (SMC) | Very High | Low–Medium | Medium | High | Enables collaborative analytics in finance and healthcare |
| Differential Privacy (DP) | High (tunable via ε) | High | High | Medium–High (depends on privacy budget) | Suitable for large-scale machine learning tasks |
| Federated Learning (FL) | Medium–High | High | Very High | High | Well-suited for IoT and edge computing; strengthened by DP/HE |
| Hybrid Approaches | Very High | Medium | Medium–High | High | Combine cryptography and ML for balanced privacy-utility trade-offs |

### Challenges and Open Research Directions

Here are some challenges and open research directions are discussed below;

- **Scalability:** Many cryptographic and privacy-preserving techniques remain computationally intensive, making it challenging to process large-scale data efficiently [27]. Scaling these methods to real-time or big data applications is still an active research problem.
- **Privacy–Utility Trade-off:** Reduced data utility or accuracy is a common trade-off for strong privacy protection. The need to develop strategies to strike a balance between these opposing goals is still an issue of great concern in different fields of endeavour like in healthcare and finance.
- **Standardization and Interoperability:** Lack of a generally agreed set of standards on privacy-preserving systems restricts their use on heterogeneous platforms [28]. There are also interoperability problems between various frameworks that make integration in a multi-organization environment even harder.
- **Adversarial Threats:** Privacy systems can be attacked in ways like model inversion, membership inference, and data poisoning. One of the research concerns is to design strong defences to make them reliable and trustworthy.

Integration with Cloud Environments: Privacy-preserving methods deployed in the cloud infrastructure have security, efficiency, and transparency challenges. Coherent and safe integration is essential with an increasing amount of data processing being transferred to cloud systems

Literature Review

This section compares the literature of privacy-preserving approaches in federated and collaborative learning, which offer various mechanisms including gradient scarification, homomorphic encryption, blockchain, transfer learning, and outlier detection to optimize privacy, scalability, and efficiency, as presented in Table III.

Feng and Du (2021) FLZip is a new framework that helps to alleviate the overhead of HE in FL. Based on the size of the gradient at each layer, clients filter out non-significant gradient values instead of encrypting individual gradient values. A key-value pair encoding is employed by FLZip to aggregate the ciphertexts of the scarified gradients. Additionally, an error accumulation mechanism is integrated to ensure that model accuracy remains intact despite scarification. With no loss of accuracy in the model, FLZip offers much better gains than Batch Crypt, an advanced HE-based FL framework: 6.4 times reduction in the number of encryption and decryption operations, and 13.1 times reduction in the footprint to and from the server [29].

Itokazu, Wang and Ozawa (2021) utilize a decision tree ensemble based on the so-called federated learning scheme to propose a novel method for detecting outliers in data that is assessed by many organizations. By incorporating an existing outlier detection mode known as Isolation Forest into the federated learning paradigm, this study enables the provision of additive homomorphic encryption to safeguard the data of each entity. The experimental results from several benchmark data sets demonstrate that the proposed privacy-preserving Isolation Forest (pp-forest) achieves consistent classification performance regardless of the increase in the number of organisations. This performance is essentially identical to that under a single-organization condition [30].

Passerat-Palmbach et al. (2020) provide the groundwork for a state-of-the-art blockchain-orchestrated ML system that can do FL in healthcare while protecting patient privacy, opening up new possibilities for health-related applications. With these six pillars, this plan can succeed in the years to come (a) Anonymously stored data and analytics procedures on a public blockchain that is both secure and transparent. (b) The worth that results from data/compute matching that were before considered unethical, unlawful, or unworkable. (c) To obtain guarantees, use federated learning in conjunction with advanced cryptography. (d) Homomorphic Encryption, Secure Multi-Party Computation, and AMD SEV-SNP encryption are examples of software and hardware security capabilities. (e) Rewards for data quality are incentivised using tokenised reputation systems. (f) Eliminating incorrect data obtained by mitigating model poisoning risk [31].

Zhang et al. (2020) suggested method prevents training set exposure and hence solves the privacy problem in federated learning by communicating with the server solely about gradients. Information extraction from gradients and possible server-side result falsification are two of the existing vulnerabilities that it fixes. The method achieves privacy-preserving federated learning with low communication and compute costs by utilizing Parlier homomorphic encryption and the Chinese Remainder Theorem. Integrating bilinear aggregate signature technology further guarantees that aggregated gradients are valid while keeping efficiency and accuracy high, even with verification mechanisms enabled [32].

Gao et al. (2019) provide a transfer learning strategy to handle the covariate shift in overlapping homogeneous feature spaces, connect the many feature spaces owned by various data owners, and ensure privacy in FL. propose an HFTL framework that has three possible versions: two based on secret sharing methods and one based on an end-to-end privacy-preserving multi-party learning mechanism. Experimental evidence of the HFTL's efficacy, scalability, and security on five benchmark datasets is the culmination of their work.

Subsequently, they put it into practice by predicting in-hospital mortality using the MIMIC-III dataset, which contains very sensitive patient information [33].

Zhang et al. (2018) machine learning framework that can enhance learning efficiency by sharing user data has recently gained significant interest and begun to be implemented in real-world problems. When dealing with highly sensitive user data, data leakage is a real possibility due to the nature of collaborative deep learning, which involves multiple users exchanging and interacting with data. So, a major issue arises: how to safeguard data privacy during collaborative deep learning processing. Using two stages of collaborative deep learning, the authors of this study summaries the use of privacy-preserving technologies and examine the present status of research in this area. The future trajectory and trend of this issue are ultimately discussed [34].

TABLE III. COMPARATIVE ANALYSIS OF PRIVACY-PRESERVING APPROACHES

| Authors / Year | Proposed Framework / Approach | Techniques Used | Focus Area / Problem Addressed | Key Contributions | Results / Advantages |
|---|---|---|---|---|---|
| Feng & Du (2021) | FLZip framework to reduce overhead of HE in FL | Homomorphic Encryption (HE), Gradient Sparsification, Key-value encoding, Error accumulation | High computational overhead of HE in federated learning | Efficient encryption via gradient sparsification and error handling | Reduced encryption/decryption by 6.4×/13.1×, network footprint reduced by 5.9×/12.5×, accuracy maintained |
| Itokazu, Wang & Ozawa 2021) | Privacy-preserving Isolation Forest (pp-iForest) | Decision Tree Ensemble, Federated Learning, Additive HE | Outlier detection across multiple organizations with privacy guarantees | Extends Isolation Forest into federated setting using HE | Stable classification across datasets, accuracy close to centralized setting |
| Passerat-Palmbach et al. (2020) | AI platform for FL powered on blockchain technology | Blockchain, FL, HE, SMC, Hardware cryptography (Intel SGX, AMD SEV-SNP) | Secure orchestration of FL in sensitive medical data sharing | Integrates blockchain, cryptography, and hardware for FL | Strong privacy and compute guarantees, incentivized data quality, model poisoning prevention |
| Zhang et al. (2020) | Privacy-preserving FL with gradient sharing and verification | Paillier HE, Chinese Remainder Theorem, Bilinear Aggregate Signatures | Gradient leakage and falsification threats in FL | Protects gradients and ensures correctness via cryptographic verification | Low cost, high accuracy, secure aggregation with verification |

| Gao et al. (2019) | Heterogeneous Federated Transfer Learning (HFTL) | Homomorphic Encryption, Secret Sharing, Transfer Learning | Handling heterogeneous feature spaces across parties in FL | End-to-end secure federated transfer learning framework | Secure, effective, scalable; validated on 5 datasets + real hospital mortality prediction |
|---|---|---|---|---|---|
| Zhang et al. (2018) | Collaborative Privacy Protection Intelligent Machines | Collaborative DL, Privacy-preserving methods (various) | Risks of data leakage in collaborative deep learning | Reviews methods in collaborative DL phases and highlights gaps | Describes current practices, points out potential dangers, and proposes directions for future study |

## Conclusion and Future Directions

The processing and sharing of massive amounts of data across numerous entities have made the protection of sensitive information a major problem in cloud and distributed computing settings. Secure multi-party computation, federated learning, differential privacy, homomorphic encryption, and hybrid frameworks are some of the privacy-preserving data processing approaches that make analytics, collaborative learning, and computation possible. These approaches have been applied across healthcare, finance, IoT, and industrial domains, offering a balance between security, scalability, efficiency, and accuracy. Still, there are a lot of obstacles. Difficulties in seamlessly integrating clouds, large overheads of communications and computations, privacy/data utility trade-offs, non-standardization, and susceptibility to adversarial attacks make these approaches not widely used. To ensure that privacy-sensitive systems fulfill their claim, must address these limitations. The future research directions can be seen as creation of lightweight, scalable cryptographic protocols with application to large-scale and resource-constrained settings, the design of hybrid frameworks that complementary privacy-preserving techniques, as well as the creation of adaptable mechanisms to balance privacy and utility depending on the context. Also, standard evaluation systems and standards are required to determine the effectiveness, security and interoperability of these solutions. These areas are essential in order to implement effectively, safely, and realistically privacy-preserving processing of data in the real world.

## References

1. H. Fang and Q. Qian, "Privacy Preserving Machine Learning with Homomorphic Encryption and Federated Learning," Futur. Internet, vol. 13, no. 4, p. 94, Apr. 2021, doi: 10.3390/fi13040094.
2. Y. Bai and M. Fan, "A Method to Improve the Privacy and Security for Federated Learning," 2021 IEEE 6th Int. Conf. Comput. Commun. Syst., no. April 2021, pp. 704–708, 2021, doi: 10.1109/ICCCS52626.2021.9449214.
3. Abhishek and P. Khare, "Cloud Security Challenges: Implementing Best Practices for Secure SaaS Application Development," Int. J. Curr. Eng. Technol., vol. 11, no. 06, pp. 669–676, Nov. 2021, doi: 10.14741/ijcet/v.11.6.11.
4. M. Hao, H. Li, G. Xu, S. Liu, and H. Yang, "Towards Efficient and Privacy-Preserving Federated Deep Learning," in ICC 2019 - 2019 IEEE International Conference on Communications (ICC), IEEE, May 2019, pp. 1–6. doi: 10.1109/ICC.2019.8761267.

5. P. Mell and T. Grance, "The NIST Definition of Cloud Computing," in Application Performance Management (APM) in the Digital Enterprise, Elsevier, 2017, pp. 267–269. doi: 10.1016/B978-0-12-804018-8.15003-X.
6. A. Papaioannou and K. Magoutis, "An Architecture for Evaluating Distributed Application Deployments in Multi-clouds," in 2013 IEEE 5th International Conference on Cloud Computing Technology and Science, IEEE, Dec. 2013, pp. 547–554. doi: 10.1109/CloudCom.2013.79.
7. B. Soltani, A. Ghenai, and N. Zeghib, "Towards Distributed Containerized Serverless Architecture in Multi Cloud Environment," Procedia Comput. Sci., vol. 134, pp. 121–128, 2018, doi: 10.1016/j.procs.2018.07.152.
8. V. Singh, "Lessons Learned from Large-Scale Oracle Fusion Cloud Data Migrations," Int. J. Sci. Res., vol. 10, no. 10, pp. 1662–1666, 2021.
9. Z. S. Ageed, S. R. M. Zeebaree, M. A. M. Sadeeq, R. K. Ibrahim, H. M. Shukur, and A. Alkhayyat, "Comprehensive Study of Moving from Grid and Cloud Computing Through Fog and Edge Computing towards Dew Computing," in 4th International Iraqi Conference on Engineering Technology and Their Applications, IICETA 2021, 2021. doi: 10.1109/IICETA51758.2021.9717894.
10. G. Sarraf, "Resilient Communication Protocols for Industrial IoT: Securing CyberPhysical-Systems at Scale," Int. J. Curr. Eng. Technol., vol. 11, no. 6, 2021.
11. D. Liu, Z. Yan, W. Ding, and M. Atiquzzaman, "A Survey on Secure Data Analytics in Edge Computing," IEEE Internet Things J., vol. 6, no. 3, pp. 4946–4967, Jun. 2019, doi: 10.1109/JIOT.2019.2897619.
12. S. B. V. Naga, K. C. Sunkara, S. Thangavel, and R. Sundaram, "Secure and Scalable Data Replication Strategies in Distributed Storage Networks," Int. J. AI, BigData, Comput. Manag. Stud., vol. 2, no. 2, pp. 18–27, 2021, doi: 10.63282/3050-9416.IJAIBDCMS-V2I2P103.
13. A. Acar, H. Aksu, A. S. Uluagac, and M. Conti, "A survey on homomorphic encryption schemes: Theory and implementation," ACM Comput. Surv., vol. 51, no. 4, 2018, doi: 10.1145/3214303.
14. D. Quang, B. Martini, and C. K.-K. Raymond, "The role of the adversary model in applied security research," Comput. Secur., vol. 81, pp. 156–181, Mar. 2019, doi: 10.1016/j.cose.2018.12.002.
15. V. M. L. G. Nerella, "Architecting secure, automated multi-cloud database platforms strategies for scalable compliance," Int. J. Intell. Syst. Appl. Eng., vol. 9, no. 1, pp. 128–138, 2021.
16. F. Wang et al., "Privacy-Preserving Collaborative Model Learning Scheme for E-Healthcare," IEEE Access, vol. 7, pp. 166054–166065, 2019, doi: 10.1109/ACCESS.2019.2953495.
17. J. Zhou, Y. Feng, Z. Wang, and D. Guo, "Using Secure Multi-Party Computation to Protect Privacy on a Permissioned Blockchain," Sensors, vol. 21, no. 4, p. 1540, Feb. 2021, doi: 10.3390/s21041540.
18. T. Wang, X. Zhang, J. Feng, and X. Yang, "A Comprehensive Survey on Local Differential Privacy toward Data Statistics and Analysis," Sensors, vol. 20, no. 24, p. 7030, Dec. 2020, doi: 10.3390/s20247030.
19. P. C. M. Arachchige, P. Bertok, I. Khalil, D. Liu, S. Camtepe, and M. Atiquzzaman, "Local Differential Privacy for Deep Learning," IEEE Internet Things J., vol. 7, no. 7, pp. 5827–5842, Jul. 2020, doi: 10.1109/JIOT.2019.2952146.
20. Q. Wu, K. He, and X. Chen, "Personalized Federated Learning for Intelligent IoT Applications: A Cloud-Edge Based Framework," IEEE Open J. Comput. Soc., vol. 1, pp. 35–44, 2020, doi: 10.1109/OJCS.2020.2993259.
21. Y. Chen, X. Qin, J. Wang, C. Yu, and W. Gao, "FedHealth: A Federated Transfer Learning Framework for Wearable Healthcare," IEEE Intell. Syst., vol. 35, no. 4, pp. 83–93, Jul. 2020, doi: 10.1109/MIS.2020.2988604.

22. I. Kholod et al., "Open-Source Federated Learning Frameworks for IoT: A Comparative Review and Analysis," Sensors, vol. 21, no. 1, p. 167, Dec. 2020, doi: 10.3390/s21010167.
23. Q. Yang, Y. Liu, T. Chen, and Y. Tong, "Federated Machine Learning," ACM Trans. Intell. Syst. Technol., vol. 10, no. 2, pp. 1–19, Mar. 2019, doi: 10.1145/3298981.
24. X. Gu, M. Li, and L. Xiong, "PRECAD: Privacy-Preserving and Robust Federated Learning via Crypto-Aided Differential Privacy," arXiv, 2021.
25. A. Haleem, M. Javaid, R. P. Singh, R. Suman, and S. Rab, "Blockchain technology applications in healthcare: An overview," Int. J. Intell. Networks, vol. 2, pp. 130–139, 2021, doi: 10.1016/j.ijin.2021.09.005.
26. U. Sheikh, A. Khan, B. Ahmed, A. Waheed, and A. Hameed, "Provenance Inference Techniques: Taxonomy, comparative analysis and design challenges," J. Netw. Comput. Appl., vol. 110, pp. 11–26, May 2018, doi: 10.1016/j.jnca.2018.03.004.
27. M. A. Ferrag, L. Shu, X. Yang, A. Derhab, and L. Maglaras, "Security and Privacy for Green IoT-Based Agriculture: Review, Blockchain Solutions, and Challenges," IEEE Access, vol. 8, pp. 32031–32053, 2020, doi: 10.1109/ACCESS.2020.2973178.
28. M. Gupta, M. Abdelsalam, S. Khorsandroo, and S. Mittal, "Security and Privacy in Smart Farming: Challenges and Opportunities," IEEE Access, vol. 8, pp. 34564–34584, 2020, doi: 10.1109/ACCESS.2020.2975142.
29. X. Feng and H. Du, "FLZip: An Efficient and Privacy-Preserving Framework for Cross-Silo Federated Learning," in 2021 IEEE International Conferences on Internet of Things (iThings) and IEEE Green Computing & Communications (GreenCom) and IEEE Cyber, Physical & Social Computing (CPSCom) and IEEE Smart Data (SmartData) and IEEE Congress on Cybermatics (Cybermatics), Dec. 2021, pp. 209–216. doi: 10.1109/iThings-GreenCom-CPSCom-SmartData-Cybermatics53846.2021.00044.
30. K. Itokazu, L. Wang, and S. Ozawa, "Outlier Detection by Privacy-Preserving Ensemble Decision Tree U sing Homomorphic Encryption," in Proceedings of the International Joint Conference on Neural Networks, 2021. doi: 10.1109/IJCNN52387.2021.9534464.
31. J. Passerat-Palmbach et al., "Blockchain-orchestrated machine learning for privacy preserving federated learning in electronic health data," in 2020 IEEE International Conference on Blockchain (Blockchain), IEEE, Nov. 2020, pp. 550–555. doi: 10.1109/Blockchain50366.2020.00080.
32. X. Zhang, A. Fu, H. Wang, C. Zhou, and Z. Chen, "A Privacy-Preserving and Verifiable Federated Learning Scheme," in ICC 2020 - 2020 IEEE International Conference on Communications (ICC), IEEE, Jun. 2020, pp. 1–6. doi: 10.1109/ICC40277.2020.9148628.
33. D. Gao, Y. Liu, A. Huang, C. Ju, H. Yu, and Q. Yang, "Privacy-preserving Heterogeneous Federated Transfer Learning," in Proceedings - 2019 IEEE International Conference on Big Data, Big Data 2019, 2019. doi: 10.1109/BigData47090.2019.9005992.
34. D. Zhang, X. Chen, D. Wang, and J. Shi, "A survey on collaborative deep learning and privacy-preserving," in Proceedings - 2018 IEEE 3rd International Conference on Data Science in Cyberspace, DSC 2018, 2018. doi: 10.1109/DSC.2018.00104.